\documentclass{aa}
\usepackage{graphicx}
\usepackage[]{natbib}
\bibpunct{(}{)}{;}{a}{}{,}
\def\gsim{~\rlap{$>$}{\lower 1.0ex\hbox{$\sim$}}}
\def\lsim{~\rlap{$<$}{\lower 1.0ex\hbox{$\sim$}}}
\begin{document}
\title{Infrared spectroscopy of carbon monoxide in V838 Mon\\
during 2002--2006\thanks{Based on data obtained at the United Kingdom
Infrared Telescope, which is operated by the Joint Astronomy Center on
behalf of the U. K. Particle Physics and Astronomy Research Council.}
}
   \author{T. R. Geballe
          \inst{1}
          \and
          M. T. Rushton \inst{2}
          \and
          S. P. S. Eyres \inst{2}
          \and
          A. Evans \inst{3}
          \and
          J. Th. van Loon \inst{3}
          \and
          B. Smalley \inst{3}
          }
   \offprints{T. R. Geballe}
   \institute{Gemini Observatory, 670 N. A'ohoku Place, Hilo, HI 96720
U.S.A.\\
              \email{tgeballe@gemini.edu}\\
         \and
Centre for Astrophysics, University of Central Lancashire,
Preston, PR1 2HE, U.K.\\
\and
Astrophysics Group, Keele University, Keele, Staffordshire, ST5
5BG, U.K.}

   \date{accepted on 2007 February 23 for publication in Astronomy and
Astrophysics}

   \abstract 
{} 
{We report spectra of the overtone and fundamental
bands of CO in the eruptive variable V838~Mon, which trace the
recent evolution of the star and allow its ejecta to be characterized.} 
{The data were obtained at the United Kingdom Infrared Telescope on
fourteen nights from 2002 January, shortly after the first outburst of
the star, to 2006 April.} 
{Although the near-infrared stellar spectrum superficially resembled a
cool supergiant after both the first and third of its outbursts in 2002,
its infrared ``photosphere" at that time consisted of highly blueshifted
gas that was moving  outward from the original stellar
surface. A spectrum obtained during the third outburst reveals a
remarkable combination of emission and absorption in the CO first overtone
bands. The most recent observations show a composite spectrum that
includes a stellar-like photosphere at a temperature similar to that seen
just after the initial outburst, but at a radial velocity redshifted by 15
km~s$^{-1}$ relative to the stellar velocity determined from SiO maser
emission, suggesting that the atmosphere is now contracting. Three shell
components, corresponding to expansion velocities of 15, 85, and 145
km~s$^{-1}$, also are present, but absorption is seen at all
expansion velocities out to 200~km~s$^{-1}$.  Weak absorption features of
fundamental band lines of $^{13}$CO have been detected. However, the large
uncertainty in the value of $^{12}$C/$^{13}$C does not constrain the
evolutionary status of the progenitor.} 
{}



\maketitle

\section{Introduction}

V838 Monocerotis literally burst upon the scene in early 2002 January,
when its visual brightness increased by nearly two orders of magnitude
over its pre-eruption value, $V\sim$16~mag \citep{bro02}. Two additional
outbursts occurred, in 2002 February and March \citep[see][and references
therein for optical and infrared light curves]{mun05}. In early February
the visual brightness increased by an additional order of magnitude,
making the overall increase a factor of $\sim$5000. During the third
outburst, in March, the star nearly regained its February peak optical
brightness. No further eruptions have occurred and after 2002 mid-March
the visual magnitude quickly declined to close to the pre-eruption value.
However, the near-infrared flux from V838 Mon, which also was very strong
after the January eruption and underwent large fluctuations during the
subsequent outbursts in the first few months of 2002, has only declined
slowly since then. The optical and infrared behavior of V838 Mon, coupled
with the presence of P Cygni and (later) inverse P Cygni profiles
\citep{zwi02,geb02,kip04,rus05a} suggest that considerable material was
ejected from the star, but that some of the material was being returned.

A prominent light echo surrounding V838~Mon \citep{hem02} appeared shortly
after the second outburst. Polarimetry of it has led to an estimated
distance to V838 Mon of 5.9~kpc \citep{spa06}.  The fading optical spectrum
of V838~Mon, which resembled that of a late type star during the early
months of 2002 \citep[e.g.,][]{zwi02, kip04} eventually revealed the
presence of a B3 V component \citep{des02,wag02,mun02}, which is very
likely a companion. \citet{afs07} have classified three nearby stars, first
seen by \citet{wis03}, as B-types, and determined their mean distance to be
6.2~$\pm$~1.2~kpc, similar to that of V838~Mon, suggesting that V838~Mon
belongs to a young cluster.

The behaviors of only two other erupting variables, M31~RV \citep{ric89}
and V4332 Sgr \citep{mar99} resemble the very detailed light curve and
spectral energy distribution of V838 Mon. However, key portions of the
light curves of these objects do not exist:  the actual outburst (or
outbursts) of V4332 Sgr were not observed and observations of M31~RV were
very few and the source is now too faint to be observed. Thus the degrees
of similarity are uncertain. Various scenarios have been suggested to
account for the behaviors of these objects, invoking nova-like outbursts,
thermonuclear shell events in a massive stars, He-shell flashes in
post-asymptotic giant branch stars, and both star-planet and star-star
mergers \citep[see e.g.][and references
therein]{law05,mun05,tyl05,ret06,tyl06}.

This paper reports infrared spectra of the fundamental and first overtone
vibration-rotation bands of carbon monoxide (CO) in V838~Mon, obtained at
the United Kingdom 3.8~m Infrared Telescope (UKIRT) on Mauna Kea, from
shortly after the initial outburst of the object to 2006. We use the CO
spectra to follow the evolution of the star and its ejecta during and
after its three outbursts in 2002 and to attempt to constrain some of the
basic properties of the progenitor, including the value of
$^{12}$C/$^{13}$C.

The first vibrational overtone ($\Delta$$v$=2; where $v$ is the
vibrational quantum number) bands of CO at 2.3--2.5~$\mu$m are among the
the most prominent and characteristic spectral features of cool
($T_{eff}$$\lsim$5000~K) stars of all luminosity classes. Typical stellar
spectra containing these bands are shown in \citet{wal97}. At low-medium
resolution the stellar CO absorption breaks up into a number of individual
asymmetric and partially overlapping features corresponding to 2-0, 3-1,
4-2, ... vibrational transitions, shifted to successively longer
wavelength. The bands of different isotopic species also are shifted in
wavelength, with heavier species shifted longward.  Each feature consists
of a multitude of individual vibration-rotation lines that can be
resolved from one another at high spectral resolution.

In cool stellar photospheres a sharp edge, or band head, occurs at the
short wavelength limit of each vibrational feature, due to the coincidence
of many vibration-rotation lines. For $^{12}$C$^{16}$O the 2-0 band
head is at 2.2935~$\mu$m and marks the short wavelength onset of CO
absorption in the K band. Each $\Delta$$v$=2 band head occurs near
rotational quantum number $J$=50.  Because the excitation temperature of
this rotational level is $\sim$7000~K, $\Delta$$v$=2 band heads can be
prominent only if the absorbing gas has T$\gsim$1500~K). At lower
temperatures only absorption lines from lower $J$ levels will be present.
High densities ($n$~$>$~10$^{8}$~cm$^{-3}$ for $v$=0 and
$n$~$>$~10$^{10}$~cm$^{-3}$ for $v$$>$0) also are required for the
ro-vibrational levels producing the band heads to be populated; normally
these conditions are only found in or very close to stellar photospheres.  
The CO bands usually are seen in absorption against the stellar continuum;  
however, if the sufficiently high temperatures and densities exist outside
the photosphere the bands can be in emission.

In low-medium resolution spectra of cool stellar photospheres the
fundamental ($\Delta$$v$=1) band of CO is apparent as a broad and fairly
structureless depression from $\sim$4.5~$\mu$m to $\sim$5.0~$\mu$m. The
band heads of the fundamental occur near $J$=90; their wavelengths
(4.3--4.5~$\mu$m for the several lowest vibrational levels) are
inaccessible from ground-based telescopes and in any event the lines
forming them are very weak owing to the high rotational excitation
($\sim$22,000~K).  Einstein $A$ values of fundamental band transitions are
10--100 times larger than those of first overtone transitions from the
same levels. Hence it is possible to detect much lower column densities of
CO by observing fundamental band lines than by observing overtone band
lines from the same ro-vibrational levels. This is particularly important
when observing cool circumstellar or interstellar CO.


\section{Observations} 

Spectroscopy of CO in V838~Mon began at UKIRT on 2002 January 12 and has
continued into 2006, with data obtained on fourteen separate nights
(see Table~1). Most of the spectra were acquired in the 2.0--2.5~$\mu$m
band using the 40 lines/mm grating of the facility 2-dimensional array
spectrograph, CGS4 \citep{mou90}, or the HK grism in the
imager/spectrograph UIST \citep{ram04}, at resolving powers of R$\sim$850.
Some spectra of specific narrow wavelength intervals were obtained at
R$\sim$18,000 using CGS4's echelle. Some of the data has been reported
previously by \citet{eva03} and \citet{rus05b}. Other observers have also
obtained low resolution spectra covering CO bands during this period.  
\citep{ban02,lyn04,ban05}. In particular, \citet{ban02} and \citet{ban05}
have reported the detection of both the first and second ($\Delta$$v$=3)
overtone bands of CO in their spectra obtained during 2002--2004. We have
been unable to identify the latter bands, which are located near
1.6~$\mu$m, in our UKIRT data. We note that the weak absorption features
they identify as CO, although corresponding in wavelength to some of the
second overtone CO bands do not have the characteristic sharp short
wavelength edges (due to band head formation) and the overall asymmetries 
of CO vibrational bands. Because of the uncertainty as to the presence of
these bands, and their weakness if they are present, we restrict this
study of CO to its first overtone and fundamental bands.

\begin{table}
\begin{minipage}[t]{\columnwidth}
\centering
\renewcommand{\footnoterule}{}
\caption{CO observing log\label{table1}}
\begin{tabular}{l l l}
\hline\hline
UT Date & Resolution & Band\\
\hline
2002 Jan 12 & Low\footnote{resolution $\approx$ 350 km s$^{-1}$,
accuracy $\approx$ 20 km s$^{-1}$} & $\Delta$$v$=2\\
2002 Mar 09 & Low & $\Delta$$v$=2\\
2002 Oct 29 & Low and High\footnote{resolution $\approx$ 17 km s$^{-1}$,
accuracy $\approx$ 3 km s$^{-1}$} & $\Delta$$v$=2\\
2002 Dec 17 & Low & $\Delta$$v$=2\\   
2003 Feb 05 & High & $\Delta$$v$=1,2\\
2003 Apr 04 & Low & $\Delta$$v$=2\\ 
2003 Oct 06 & Low & $\Delta$$v$=2\\ 
2003 Dec 06 & High & $\Delta$$v$=2\\
2005 Jan 05 & Low & $\Delta$$v$=2\\ 
2005 Feb 25 & High & $\Delta$$v$=2\\
2005 Mar 02 & High & $\Delta$$v$=1\\
2005 Mar 04 & High & $\Delta$$v$=1\\
2006 Feb 13 & Low & $\Delta$$v$=2\\ 
2006 Apr 18 & High & $\Delta$$v$=2\\
\hline
\end{tabular} 
\end{minipage}
\end{table}

All of the UKIRT spectra reported here were taken in the conventional
stare/nod-along-slit mode.  Extracted spectra were flat-fielded,
de-spiked, and wavelength-calibrated using telluric absorption lines or
emission lines from arc lamps.  Accuracies in absolute wavelength are
better than 0.00017~$\mu$m (20~km~s$^{-1}$) for the low resolution
spectra and 3~km~s$^{-1}$ for the echelle spectra. Spectra of early type
standard stars were obtained nearly simultaneously with and close to the
same airmass as the spectra of V838~Mon, in order to accurately remove
telluric features.

\section{Results and Discussion}

Here we provide descriptions of the most significant spectra of CO
obtained during 2002--2006 and use them to derive velocity information,
temperatures, and physical locations of the CO, and to constrain the
$^{12}$C/$^{13}$C ratio. We attempt to physically relate the different
spectral features and velocity components to the recent evolution of the
source.  A subsequent paper (Rushton et al., in preparation) will present
detailed modeling of some of these spectra.

Because of the high speeds of the ejected gas, even at the low resolution
of these spectra, the accuracy of the wavelength calibration is sufficient
for using the band heads to constrain the locations of the principal
line-forming regions, by comparison of wavelengths at which the observed
band heads occur to the wavelengths expected for gas at the stellar
velocity. In particular, the 2-0 band head is an accurate velocity
indicator, unless it is extremely saturated, which should not be the case
here.

   \begin{figure}
   \resizebox{\hsize}{!}{\includegraphics{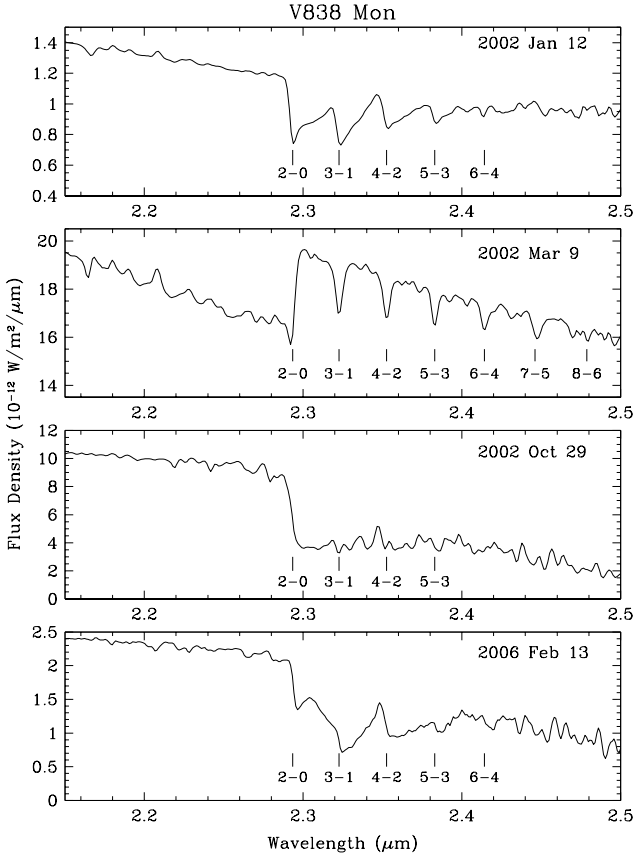}}
\caption{Low resolution spectra of the first overtone bands of CO in
V838~Mon. The positions of band heads of $^{12}$CO are indicated.}
    \end{figure}
   
   \begin{figure}
      \resizebox{\hsize}{!}{\includegraphics{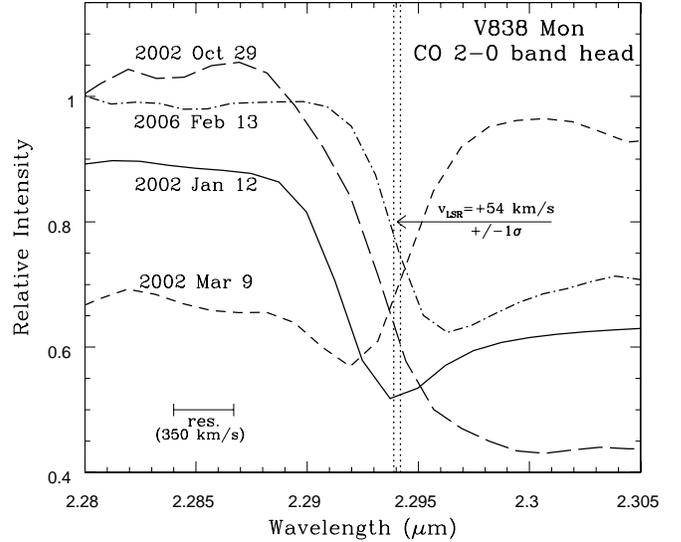}}
\caption{Details of the low resolution spectra in Fig.~1 near the CO 2-0 
band head. The dashed vertical lines are 20~km~s$^{-1}$ (1$\sigma$) to 
either side of the wavelength that the band head would appear for CO at
the stellar radial velocity, and would bisect the absorption edge if
the band head were formed at that velocity.}
    \end{figure}

Based on measurements of two SiO maser transitions at millimeter
wavelengths, \citet{deg05} has determined the radial velocity of V838~Mon
to be +54~km~s$^{-1}$ LSR (+71~km~s$^{-1}$ heliocentric; LSR velocities
are used hereafter). Their result assumes that amplification tangential to
the circumstellar material dominates the emission, which is generally true
in late-type stars \citep[e.g.,][]{jia95}, but we note that V838 Mon is a
highly unusual case, and also that it may be in a binary system.

\subsection{Low resolution spectroscopy of the $\Delta$$v$=2 bands}

Figure~1 shows low-resolution 2.15--2.50~$\mu$m spectra of V838~Mon
obtained on three dates in 2002 and one date in 2006. Figure~2 shows
details of the same spectra near the wavelength of the 2-0 band head of
$^{12}$CO.  

The spectrum in the top panel of Fig.~1 was obtained on 2002 January 12,
roughly two weeks after the initial outburst. Superficially it looks
similar to that of a stellar photosphere. However, a detailed examination
of the spectrum near the 2-0 band head (see Fig.~2) reveals that the CO
absorption edge is highly blueshifted relative to the radial velocity of
the star. This is also the case for the 3-1, 4-2, and 5-3 band heads.
Thus, the hot CO contributing to these absorption bands is in dense gas
that was moving outward at speeds of up to a few hundred km~s$^{-1}$.

The second spectrum in Fig. 1, obtained on 2002 March 9, during the
second outburst, may be the most unusual example of CO first overtone
band emission ever reported. When in emission the $\Delta$$v$=2 bands
commonly have peaks near their heads, and decreasing emission to longer
wavelengths \citep[e.g., see][]{cha93}. The 2002 March 9 emission does
not resemble this in any respect. Assuming that the continuum level may
be extrapolated from its slope at 2.15--2.29~$\mu$m, the CO line emission
longward of each band head is nearly constant with wavelength. 
Narrow absorption features are located at the wavelengths of the band
heads, absorbing out much of the emission there. To our knowledge neither
the constancy of the emission between the band heads nor the absorption
dips at the band heads have been observed previously in an astronomical
object. This remarkable CO emission did not last long, as the spectrum of
\citet{ban02} obtained two weeks later shows only CO absorption.

Figure~2 indicates that on 2002 March 9 the maximum absorption at
the 2-0 band head was shifted by more than $-$200~km~s$^{-1}$ relative to
the stellar velocity. However, the low resolution of the spectrum and the
presence of strong emission on only the long wavelength side of the
absorption conspire to cause the observed absorption maximum to be
blueshifted from the actual velocity of peak absorption. The peak
absorptions at each of the other band heads, where the CO emission on
each side is of comparable strength, are blueshifted by
80~$\pm$20 km~s$^{-1}$, which we take as a more reliable value of the
velocity shift relative to the star. 

Between the band heads the spectrum of 2002 March 9 must be a
superposition of CO emission and absorption, because it is physically
impossible for absorption to be present only near the band heads at
$J$$\sim$50 and not at other values of $J$. If both the emission and
absorption lines were optically thin, in most cases the composite
spectrum, as viewed at low resolution, would approach closer to the level
of the (extrapolated) continuum further from the band head, because the
increasing line spacing with increasing wavelength across each band leads
to the both the emission and absorption per unit wavelength decreasing
steadily. That behavior at also would be observed even if the CO emission
lines were optically thick, as long as the lines were narrow, because of
the increasing line spacing. Separately CO absorption and emission
normally behave in this manner; see \citet{wal97} and \citet{cha93}. Thus,
the dominance of emission over absorption and especially the ''flatness"
of the emission between the band heads indicate that on 2002 March 9 the
CO emission lines were optically thick over a sufficiently wide range of
velocities ($\sim$200~km~s$^{-1}$) to fill the gaps between lines far from
the band heads. The similar slopes of the CO emission and the continuum
shortward of the 2-0 band head and the $\sim$20\% excess of the emission
over the continuum indicates that the two were emitted by gas at similar
temperatures and that the CO radius was roughly 10\% larger than the
stellar radius at that time.

Because on 2002 March 9 CO absorption was present up to at least the
$v$=7-5 band, the gas producing it was quite dense and was at least as hot
as the CO observed in 2002 January. Yet, if the emission was optically
thick, as we believe, the absorption features must have formed outside of
the emitting region. The absorbing CO may have been located on the outer
surface of the gas producing the emission. Alternatively, it may have been
ejected earlier than 2002 March 9, for example during the major outburst
in early 2002 February. Unfortunately no CO data were obtained precisely
at the time of the February outburst; however, roughly one week later
\citet{hin02} detected broad and shallow CO absorption lines blueshifted
by 80 km~s$^{-1}$ relative to the (now known) heliocentric velocity, thus
approximately the velocity of the CO absorption features that we observed
on March~9. Nevertheless, it is difficult to understand how such a
separate hot and dense circumstellar component could persist for even one
month, and therefore we believe that the CO line emission and absorption
on March~9 occurred in close physical proximity to one another.

The third spectrum in Fig.~1 was obtained in 2002 October, approximately
ten months after the initial eruption. In strong contrast to the previous
spectrum, it shows what may be the deepest CO overtone absorption reported
in a star, as well as a remarkably broad 2-0 band head absorption edge. The
great depth is indicative of the presence of a large column density of cool
absorbing gas.  Model spectra by \citet{rus06} suggest a mean CO
temperature of 1300~K.  The observed spectrum contains only marginal
indications of the 3-1, 4-2, and 5-3 band heads of CO, even though from the
overall breadth of the absorption it is likely that those bands contribute
and thus that a somewhat warmer gas component also was present. In an
oxygen-rich star such low temperatures imply that in addition to CO lines
of H$_{2}$O constituted a significant part of the absorption beyond
2.3~$\mu$m. For V838 Mon this inference is supported by the even deeper
absorption bands of H$_{2}$O at 1.1, 1.4, and 1.9~$\mu$m in the full
1--2.5$\mu$m spectrum from 2002 October \citep{eva03}, which together with
the strong CO first overtone band led \citet{eva03} to refer to the object
as an ``L supergiant." The CO 2-0 absorption edge in the 2002 October
spectrum shown in Fig.~2 is considerably broader than in the other spectra
in the figure and is much broader than the resolution; it indicates that
CO absorbing over a wide range of velocities (perhaps 600~km~s$^{-1}$) as
well as at a wide range of temperatures contributed to the spectrum at that
time.

From 2002 October to the date of the most recent low-resolution spectrum
(an interval of 3.5 years) the CO $\Delta$$v$=2 absorption has changed
much more slowly than it did during 2002. In the most recent
low-resolution spectrum from 2006, shown at the bottom of Fig.~1, at least
four bands can be seen in absorption, clearly demonstrating the presence
of dense molecular gas at temperatures typical of an M-type stellar
photosphere. It is unlikely that the gas ejected at high velocities in
2002 could still be sufficiently hot and dense to populate the
ro-vibrational levels that produce these features.  Thus, even in the
absence of radial velocity information, one would have to conclude that
the band heads are formed in the stellar photosphere or very close to it.
Radial velocity data for the CO are available, however, both at low and
high resolution. The portion of the low resolution spectrum shown in
Fig.~2 demonstrates that the velocity of the 2-0 band head is no longer
blueshifted and is in fact close to that of the star. Moreover, the
absorption edge is considerably sharper than in 2002 October (the drop in
flux density, from 10\% to 90\% of peak absorption, occurs over
$\sim$350~km~s$^{-1}$ (compared to $\sim$700~km~s$^{-1}$ earlier) and is
unresolved.  Both of these findings suggest that in 2006 the CO forming
the band heads was photospheric or near-photospheric. A cool and
presumably detached circumstellar component is still present, clearly
revealed in Fig.~1 by the excess absorption from 2.31~$\mu$m to
2.34~$\mu$m where R branch lines from low lying $J$ levels in the CO
$v$=2-0 band are located; note that the level of flux density decreases
longward of 2.31~$\mu$m and is lower at the 3-1 band head is deeper than
at the 2-0 and 4-2 band heads. The gas producing this excess is too cool
to contribute significant absorption at $J\sim$50, where the 2-0 band head
is formed. It seems likely that the cool gas was ejected at higher
velocity and is now far from the star, but spectra at much higher
resolution, some of which are presented and discussed below, are required
to clearly demonstrate this, as well as accurately measure the velocity of
the hot, band head-forming CO.

\subsection{ High resolution spectroscopy}

The first high resolution spectra of CO were obtained more than seven
months after the final outburst of V838~Mon (see Table~1).  These early
spectra are very complex, with broad and blended absorption components, and
are still being analyzed.  The CO line profiles in spectra from 2005 and
2006 are less blended, although still highly structured, and do allow some
understanding of conditions in the absorbing gas. In the following we
present and discuss only these most recent high resolution spectra.

Figure~3 shows the 2.28-2.35~$\mu$m spectrum of V838~Mon obtained in
2006 April at a resolution of 17~km~s$^{-1}$. Many individual CO lines can
be seen. The spectrum is consistent with the low resolution spectrum
obtained two months earlier and shown in the bottom panel of Fig.~1. In
particular the 2--0 and 3--1 band heads are clearly present; in Fig.~3 the
latter can be seen in the midst of a forest of lines from the 2--0 band.

Examination of Fig.~3 reveals that the low $J$ lines of the 2--0 band
have more complex profiles than the high $J$ lines, which are presumably
largely formed in or close to the photosphere. Figure~4 shows velocity
profiles of three 2--0 lines from this spectrum, R(28), R(16), and R(6),
with widely different lower state energies (corresponding to temperatures
of 2240~K, 752~K, and 116~K, respectively). Note that the velocity
coverage is less for the higher $J$ lines, because of the decreasing 
spacing of adjacent lines in the band with increasing $J$. These specific
lines were chosen to minimize contamination by other CO lines. The R(28)
line is coincident with the 2--0 R(73) line, which should be very weak.  
The R(16) line is equidistant from the extremely weak 2--0 R(84) and
R(85) lines. The R(6) line, which should be negligibly contaminated by
the nearly coincident 2--0 R(95) line, is coincident within a few
km~s$^{-1}$ with the 3-1 R(27) line which, because its lower vibrational
level is excited, should only contribute absorption in the
warm and dense photosphere.

The high $J$ lines such as R(28) are slightly asymmetric, with blueshifted
shoulders, and are dominated by a single velocity component at $v_{\rm
LSR}=+69$~km~s$^{-1}$.  The same velocity is derived from the wavelength of
the 2--0 band head at $J$=51. One would normally expect these lines to
arise in the photosphere of the star. However, they are redshifted by
15~km~s$^{-1}$ with respect to the velocity of peak SiO maser emission,
which we have taken (with the aforementioned caveats) to be the stellar
radial velocity. This suggests that gas cotaining this CO is falling back
onto the star or the star's photosphere is contracting to approximately its
pre-outburst radius. If so the velocity of this component and possibly also
the strength of the CO overtone absorption could change noticeably in the
next few years.  Eventual contraction of the envelope is predicted by
\citet{tyl06} following the merger of a low mass star with an 8~M$_{\sun}$
main sequence star.

\begin{figure}
      \resizebox{\hsize}{!}{\includegraphics{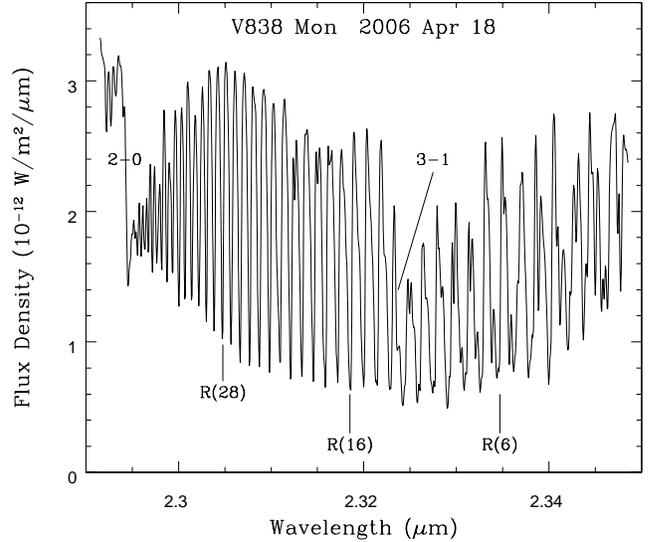}}
\caption{High-resolution spectra of first overtone bands of CO.  The    
locations of the 2-0 and 3-1 band heads of $^{12}$CO are indicated. The 
three lines of CO, whose velocity profiles are shown in Fig.~4, are
indicated.}
    \end{figure}

\begin{figure}
   \resizebox{\hsize}{!}{\includegraphics{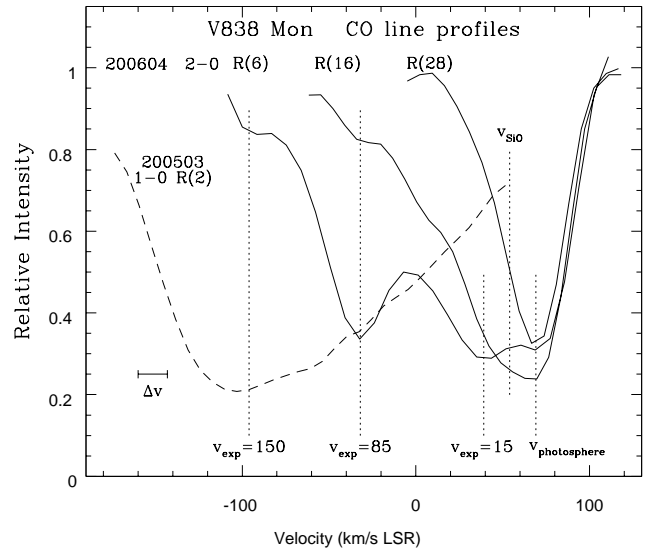}}
\caption{Velocity profiles of three 2-0 band lines and one 1-0 band
line of CO. Velocities of maximum absorption are indicated by vertical  
dashed lines, as is the stellar radial velocity obtained from SiO maser
measurements by  \citet{deg05}. Expansion velocities (relative to SiO) of
the cool CO components are given.}
    \end{figure}  

The peak absorption of the R(16) velocity profile is 
at the same radial velocity as that of the R(28) profile, but the blue
asymmetry of the R(16) profile is much more pronounced. In
addition, a weak absorption component is present at $-$32~km~s$^{-1}$
(shifted by $-85$~km~s$^{-1}$ from the nominal stellar velocity). In the
R(6) profile the $-85$~km~s$^{-1}$ component is much stronger. In
addition, the R(6) profile contains a deep absorption at +38~km$^{-1}$,
corresponding to an expansion velocity of 15 km~s$^{-1}$. It is
barely resolved from the photospheric component at +69~km~s$^{-1}$. The
blue asymmetries of the higher $J$ lines are probably caused by the
+38~km~s$^{-1}$ component, which is weaker at higher $J$, but spectral
modeling at high resolution is required to confirm this.

The 2--0 R(6) profile also contains evidence for a third and even more
blueshifted component at approximately $-96$~km~s$^{-1}$.  A
spectrum of a portion of the fundamental (1--0) band of CO near
4.7~$\mu$m, shown in Fig.~5, obtained a year earlier, demonstrates that
this component is real and that it is the dominant absorber in the
$\Delta$$v$=1 band. The spectrum is highly complex and likely
contains numerous lines of H$_{2}$O in addition to CO. However, the
strongest lines by far are the CO lines. The peak absorptions in these
lines are uncertain because the continuum level is unknown, but based on
Fig.~5 they must be at least 80\%.

Figure~4 includes the velocity profile of the 1--0 R(2) line, one of the
least contaminated of these fundamental band lines, in addition to the
profiles of the three 2--0 band lines disussed previously. The profile is
remarkably broad and asymmetric, characteristics that are present in all
of the $^{12}$CO 1--0 lines in Fig.~5. The CO at the velocity of maximum
absorption is moving outward from V838~Mon at roughly 150~km~s$^{-1}$ and
absorption is present at least to expansion velocities of 200~km~s$^{-1}$.
Information at higher velocities is unavailable because of contamination
from other lines. The decrease in absorption toward lower expansion
velocites is smooth and more gradual, with little or no evidence for the
discrete velocity components seen in the 2--0 band line profiles. The
differences between the fundamental and overtone indicate that the
continuum-forming regions at 2.3~$\mu$m and 4.7$\mu$m are located at
different radii. In particular, much of the gas producing the prominent
absorption features in the 2--0 band lines at expansion velocities of 15
and 85 km~s$^{-1}$ must be located inside the 4.7~$\mu$m continuum-forming
region. The broad 1--0 profile also demonstrates that gas at a remarkably
wide range of expansion velocities lies outside of the 4.7~$\mu$m
continuum-forming surface, far from the star.

Thus, four discrete velocity components are identified in these CO
spectra, three of which are in ejected gas. Rough excitation temperatures
can be estimated for all but the most blueshifted component and CO column
densities can be estimated for the +38 and $-$32 km~s$^{-1}$ components.  
From the relative strengths of the $\Delta$$v$=2 band heads in synthetic
spectra \citep{rus06}, the temperature of the photospheric (+69
km~s$^{-1}$) component was $\sim$2000~K in 2002 January, but has been
difficult to determine since then due to contamination from the other
components. At +38~km~s$^{-1}$ and $-$32~km~s$^{-1}$ the relative
strengths of the three overtone lines shown in Fig.~4 imply excitation
temperatures ($T$$_{ex}$) of $\sim$500~K and $\sim$200~K, respectively.  
The kinetic temperatures could be significantly higher, particularly in
the case of the cooler ($-32$~km~s$^{-1}$) component. Both the low
temperatures and the expansion velocities imply that by 2006 the
blueshifted absorption components were far from the star. The distances
that the gas producing them would have traveled from the star since early
2002 at the measured velocities are 12 A.U., 70 A.U., and 120 A.U. In
view of its low expansion velocity and the probability of significant
deceleration during the four years since it was ejected, the innermost
(+38~km~s$^{-1}$) component may have been considerably more distant than
12 A.U. from V838~Mon in 2006.

Rough estimates of column densities can be derived from CO overtone spectra
for those shell components whose temperatures are known. We find $N$$_{\rm
CO}$ in each of the +38 (500~K) and $-$32~km~s$^{-1}$ (200~K) components is
$\sim$1$\times$10$^{19}$~cm$^{-2}$, indicating hydrogen column densities of
$\sim$4~$\times$~10$^{22}$~cm$^{-2}$ if C/H is normal.  The extinction
normally associated with such column densities, $\approx$~20~mag in each
shell, is more than an order of magnitude greater than that observed
\citep{afs07}, indicating that either little dust has formed as a result of
the outbursts or that dust-forming elements are severely depleted relative
to carbon. For normal C/H and assuming spherical symmetry and the above
radii for the shells, the masses associated with the +38~km~s$^{-1}$ and
$-$32~km~s$^{-1}$ components are $\sim$1~$\times$~10$^{-5}$~M$_{\sun}$ and
$\sim$4~$\times$~10$^{-4}$~M$_{\sun}$, respectively.

\begin{figure}
    \resizebox{\hsize}{!}{\includegraphics{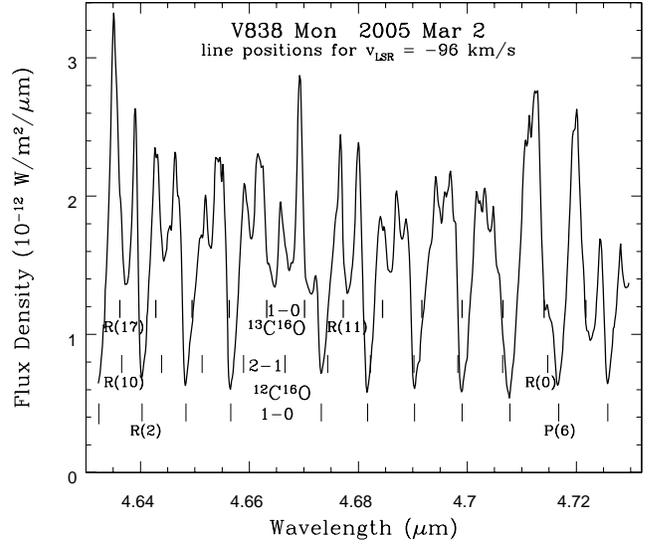}}
\caption{High resolution spectrum of a portion of the CO fundamental, 
showing strong lines near the center of the 1-0 band of $^{12}$CO.  Line 
positions are indicated at the bottom of the figure. Also indicated are 
line positions of the 2-1 band of $^{12}$CO and the 1-0 band of 
$^{13}$CO.}
    \end{figure}

\subsection{$^{12}$C/$^{13}$C}

The $^{12}$C/$^{13}$C ratio is a diagnostic of stellar evolution in
oxygen-rich stars, with giants and supergiants having ratios far below the
solar value (89) or typical interstellar values of 60--100
\citep[e.g.,][]{mer79}. Currently V838~Mon has the luminosity and
temperature of a cool giant or supergiant.  If it were a giant or
supergiant prior to its outburst, and if the bulk of the ejected gas
belonged to it (rather than to a merged companion) prior to the outburst,
one would expect that its value of $^{12}$C/$^{13}$C would be low compared
to the above values.

The 2-0 and 3-1 band heads of $^{13}$C$^{16}$O, at 2.345~$\mu$m and
2.374~$\mu$m, respectively, are not present in any of the low resolution
spectra in Fig.~1 or any of the other K-band spectra we have
obtained.\footnote{\citet{rus05b} reported the detection of weak absorption
bands of $^{13}$CO in the 2002 March 9 spectrum. We have re-examined this
spectrum and conclude that the wavelengths of the weak features do not
match those of $^{13}$CO} In most late-type stars these band heads are
readily apparent, even when $^{12}$C/$^{13}$C~$\sim$~30.  Comparisons of
some of these spectra with simple model overtone spectra \citep{rus06} have
also failed to identify lines of $^{13}$CO. Realistic synthetic spectra are
needed, but are difficult to construct due to the complexities of the star
and its multiple ejected gas shells. 

The spectrum of part of the fundamental band shown in Fig.~5 covers the
wavelengths of a number of individual lines of $^{13}$CO from its 1--0
band. The $^{13}$CO 1--0 transitions from the lower $J$ levels in this
spectral interval are the same ones that are prominent in $^{12}$CO.
Because the spectrum is badly contaminated by lines of H$_{2}$O, some of
which approach the strengths of the $^{12}$CO lines, only the $^{12}$CO
1--0 lines and the water lines are readily identified. Nevertheless, from
detailed examination of the spectrum at the $^{13}$CO line wavelengths
which do not coincide with strong $^{12}$CO or H$_{2}$O lines, we believe
we detect the presence of $^{13}$CO in absorption at
$v$$_{LSR}$~$\approx$~$-96$~km~s$^{-1}$, the velocity of peak $^{12}$CO
absorption. The absorptions are $\sim$10\% below the average of the flux
levels at adjacent wavelengths, implying that the optical depths at the
centers of these absorption lines are at least 0.1.

It is not straightforward to determine a reliable value of
$^{12}$C/$^{13}$C from these data. The best opportunity for a constraint on
the ratio comes from comparison of the fundamental band lines of $^{13}$CO
and -96~km~s$^{-1}$ components of the overtone lines of $^{12}$CO. From
Fig.~4, the peak optical depth of the $-96$~km~s$^{-1}$ component of the
2--0 R(6) line of $^{12}$CO was $\sim$0.15 in 2006 April. It would
have been somewhat greater than that a year previous when the fundamental
band lines of $^{13}$CO shown in Fig.~5 were observed. Assuming that this
component is formed completely outside of both the 2.3~$\mu$m and
4.7~$\mu$m continuum surfaces, the peak optical depths of the
$^{12}$C$^{16}$O 1--0 band lines from $J$$\sim$6 in Fig.~5 should be
$\sim$10. Thus, if $^{12}$C/$^{13}$C were 10 or less, absorption lines of
the $^{13}$C$^{16}$O 1--0 band from $J$$\sim$6 probably would have optical
depths near unity and ought to have been seen easily in the spectrum of the
fundamental band; that they are not suggests that
$^{12}$C/$^{13}$C~$\sim$100. However, in view of the complexity of the
spectrum, we feel that values as low as 10 cannot be excluded.

Due to the large uncertainty in $^{12}$C/$^{13}$C we cannot rule out any
of the proposed scenarios for the outburst of V838~Mon. We note that in
the only late He-shell flash object where this isotopic ratio is measured,
Sakurai's Object, the ratio is approximately 4 \citep{pav04,asp99}, a
value that is inconsistent with the V838~Mon spectra. However, it is not
known if such a low value applies to all cases of stars undergoing late
thermal pulses. The stellar merger model \citep{tyl06}, which invokes main
sequence stars, predicts a solar/interstellar value for $^{12}$C/$^{13}$C,
whereas in the giant star - planet merger model \citep{ret06}
$^{12}$C/$^{13}$C probably would be considerably lower. More stringent
observational constraints for V838~Mon should be possible from future
infrared spectroscopy, as the CO spectrum simplifies due to the cooling of
the expanding circumstellar shells and the decrease of CO column densities
in them, or possibly from measurements of pure rotational transitions of
CO arising in the cool extended envelope.

\begin{acknowledgements}

The authors thank the staff of the Joint Astronomy Centre for its support 
of the many observations on which this paper is based. T. R. G.'s research 
is supported by the Gemini Observatory, which is operated by the 
Association of Universities for Research in Astronomy, Inc., on behalf of 
the international Gemini partnership of Argentina,Australia, Brazil, 
Canada, Chile, the United Kingdom, and the United States of America. M. T. 
R.'s research is supported by the University of Central Lancashire.

\end{acknowledgements}

\end{document}